%% file: sample-manuscript.tex
\documentclass[sigconf]{acmart}
\AtBeginDocument{%
  }

\copyrightyear{2024} 
\acmYear{2024} 
\setcopyright{acmlicensed}\acmConference[DIS '24]{Designing Interactive Systems Conference}{July 1--5, 2024}{IT University of Copenhagen, Denmark}
\acmBooktitle{Designing Interactive Systems Conference (DIS '24), July 1--5, 2024, IT University of Copenhagen, Denmark}
\acmDOI{10.1145/3643834.3661517}
\acmISBN{979-8-4007-0583-0/24/07}

\newcommand{\AutoIcon}{\includegraphics[scale=0.03]{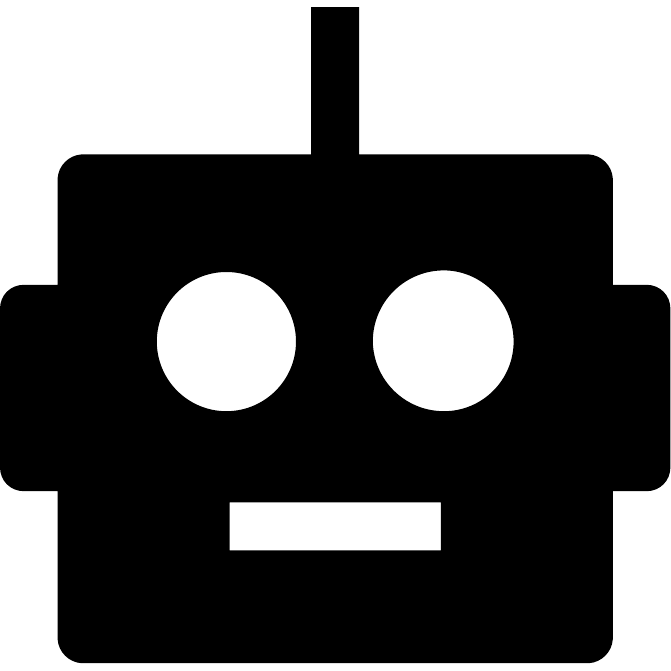}}
\newcommand{\humanIcon}{\includegraphics[scale=0.22]{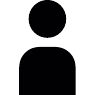}}
\usepackage{color}
\usepackage{siunitx}
\usepackage{xcolor}
\usepackage{makecell}
\usepackage{booktabs}
\usepackage{multirow}
\usepackage{graphicx}
\usepackage{hyperref}
\usepackage{csquotes}

\begin{document}

\title[Towards Feature Engineering with Human and AI's Knowledge]{Towards Feature Engineering with Human and AI's Knowledge: Understanding Data Science Practitioners' Perceptions in Human\&AI-Assisted Feature Engineering Design}


\author{Qian Zhu}
\authornote{Both authors contributed equally to the paper.}
\orcid{0000-0001-5108-3414}
\affiliation{
  \institution{The Hong Kong University of Science and Technology}
  \city{Hong Kong}
  \country{China}
}
\email{qian.zhu@connect.ust.hk}

\author{Dakuo Wang}
\authornotemark[1]
\orcid{0000-0001-9371-9441}
\affiliation{
  \institution{Northeastern University}
  \city{Boston}
  \country{United States}
}
\email{d.wang@northeastern.edu}

\author{Shuai Ma}
\orcid{0000-0002-7658-292X}
\affiliation{
  \institution{The Hong Kong University of Science and Technology}
  \city{Hong Kong}
  \country{China}
}
\email{shuai.ma@connect.ust.hk}

\author{April Yi Wang}
\orcid{0000-0001-8724-4662}
\affiliation{
  \institution{ETH Zurich}
  \city{Zurich}
  \country{Switzerland}
}
\email{april.wang@inf.ethz.ch}

\author{Zixin Chen}
\orcid{0000-0001-8507-4399}
\affiliation{
  \institution{The Hong Kong University of Science and Technology}
  \city{Hong Kong}
  \country{China}
}
\email{zchendf@connect.ust.hk}

\author{Udayan Khurana}
\orcid{0000-0001-8113-1210}
\affiliation{
  \institution{IBM Research AI}
  \city{Yorktown Heights}
  \country{United States}
}
\email{ukhurana@us.ibm.com}

\author{Xiaojuan Ma}
\orcid{0000-0002-9847-7784}
\affiliation{
  \institution{The Hong Kong University of Science and Technology}
  \city{Hong Kong}
  \country{China}
}
\email{mxj@cse.ust.hk}

\renewcommand{\shortauthors}{Qian Zhu, et al.}

\begin{abstract}
As AI technology continues to advance, the importance of human-AI collaboration becomes increasingly evident, with numerous studies exploring its potential in various fields. One vital field is data science, including feature engineering (FE), where both human ingenuity and AI capabilities play pivotal roles. 
Despite the existence of AI-generated recommendations for FE, there remains a limited understanding of how to effectively integrate and utilize humans' and AI's knowledge.
To address this gap, we design a readily-usable prototype, human\&AI-assisted FE in Jupyter notebooks. It harnesses the strengths of humans and AI to provide feature suggestions to users, seamlessly integrating these recommendations into practical workflows.
Using the prototype as a research probe, we conducted an exploratory study to gain valuable insights into data science practitioners' perceptions, usage patterns, and their potential needs when presented with feature suggestions from both humans and AI. 
Through qualitative analysis, we discovered that the \textit{``Creator''} of the feature (i.e., AI or human) significantly influences users' feature selection, and the semantic clarity of the suggested feature greatly impacts its adoption rate. Furthermore, our findings indicate that users perceive both differences and complementarity between features generated by humans and those generated by AI. Lastly, based on our study results, we derived a set of design recommendations for future human\&AI FE design.
Our findings show the collaborative potential between humans and AI in the field of FE. 
\end{abstract}





\maketitle
\input{Sections/01-introduction}
\input{Sections/02-relatedwork}

\input{Sections/03-framework}
\input{Sections/04-study}
\input{Sections/05-result}
\input{Sections/06-discussion}
\input{Sections/07-conclusion}

\begin{acks}
This research was supported in part by HKUST 30 for 30 Grant No.:  3030\_003.
\end{acks}
\bibliographystyle{ACM-Reference-Format}
\bibliography{sample-base}










\end{document}

%% file: Sections/01-introduction.tex
\section{Introduction} \label{intro}
With the rapid development of Artificial Intelligence (AI) technologies, we have seen examples of AI providing assistance for humans in various scenarios, ranging from daily life situations to formal work environments \cite{dilsizian2014artificial, khandani2010consumer, yang2018insurance, ma2022modeling, ma2019smarteye, ma2022glancee, ma2024beyond, ma2024you}. 
In light of this, HCI researchers have argued the urge to explore the cooperation needs of humans and AI in specific scenarios as future workplaces will rely on close cooperation between humans and AI \cite{buccinca2021trust, zhang2020effect, muller2019data, wang2019human, ma2024towards}.

While the growing amount of research on the collaboration between humans and AI \cite{gomez2023mitigating, ma2023should, dastin2018amazon, dilsizian2014artificial, lai2020chicago, cheng2019explaining, yang2018insurance}, there are still many unexplored needs in the data science (DS) workflow where AI is frequently utilized~\cite{wang2021autods,kross2021orienting}. 
Within the DS workflow, feature engineering (FE) --- the process of transforming the raw data into a set of features for the model to take as input --- is one of the most critical steps to bridge the knowledge from the data and the machine learning (ML) models~\cite{drozdal2020trust, muller2019data, khurana2018feature}. 
However, when dealing with FE, users often face challenges, such as the lack of domain-specific knowledge, the time-consuming process of deriving sufficient features from raw data, etc \cite{wang2019human, sambasivan2021everyone}.

To help DS practitioners with FE, researchers have proposed AI-assisted FE approaches, which use AI techniques to derive features from input data automatically~\cite{kaggle2020feature, kanter2015deep, lam2017one, yu2021feature, khurana2016cognito}. It takes advantage of AI in generating a huge number of candidate features quickly, while it usually functions as a ``black box'' and users may find it hard to interpret how and why specific features are generated \cite{duboue2020art, chatzimparmpas2021featureenvi}. 
Thus, many DS practitioners primarily rely on the time-consuming way to manually create features while domain knowledge is often scarce~\cite{wang2021autods, smith2020enabling}. 
Smith et al. built a collaborative tool to facilitate cooperation among users to construct and share features~\cite{smith2020enabling}. However, the tool is infeasible to be widely applied in practice as it requires the active participation and contribution of a group of DS practitioners, especially when dealing with a new task where the community is under-constructed~\cite{sambasivan2021everyone}.

Recognizing the constraints of AI-generated features and the limited availability of human-created features, we propose to \textbf{utilize and integrate human and AI resources collectively} to enhance FE, instead of relying solely on one side. 
We envision future systems for human\&AI FE that can harness the strengths of both humans and AI to achieve a synergistic effect. As a first step towards this goal, in this paper, we aim to understand users' perceptions and usage patterns when leveraging the combined knowledge of humans and AI for FE.

To achieve this goal, we first designed and implemented a prototype to integrate human and AI features collectively and embedded them into real-world DS pipelines~\cite{wang2020autoai}. 
We incorporated cutting-edge methodologies from both the AI and HCI domains to recommend AI-generated and human-generated features to users using identical input data~\cite{kanter2015deep, smith2020enabling}. 
For the user interface component, we adhered to established practices in real-world DS work by developing a tabular-based interface equipped with supportive functions. We integrated this interface as a plugin for widely-used online notebook platforms to ensure seamless integration and accessibility~\cite{April2019Notebook}. 
Then, taking the readily-usable human\&AI FE design as a research probe, we explore how DS practitioners perceive and use the human\&AI FE in practice, with the following research questions in mind.
\begin{itemize}
    \item \textbf{RQ1}: How do DS practitioners inspect and select features that are generated using human\&AI-assisted FE?
    \item \textbf{RQ2}: What are DS practitioners' perceptions and attitudes toward the human\&AI-assisted FE?
    \item \textbf{RQ3}: What aspects of a human\&AI-assisted FE recommendation system should be considered for improving its usability and user experience?
\end{itemize}

To answer these questions, we conducted a user study with 14 DS practitioners recruited from both industry and academia. Participants were asked to finish a real feature engineering task using our proposed human\&AI FE. We gave them sufficient freedom to explore, inspect, and leverage the suggested features, and we recorded their behavioral and feedback data for qualitative analysis. 
In addressing \textbf{RQ1}, our results showed that users tended to choose features from both sides and keep a balance between human- and AI-generated features. Besides, we observed varying levels of reliance on the ``\emph{Creator}'' (i.e., human or AI) and discovered that feature explainability and individuals' understanding of semantics influenced their decisions. We identified similarities and differences in feature selection behaviors and strategies.
Regarding \textbf{RQ2}, participants recognized distinctions and complementarities between human- and AI-generated features. 
They reported a positive user experience towards the proposed human\&AI FE, especially the designed user interface and the integration with a standard and usable data science workflow.
Lastly, for \textbf{RQ3}, we derived a set of practical recommendations to enhance future human\&AI FE design, based on the analysis of participants' behaviors and feedback.

Our work contributes new knowledge on understanding how DS practitioners perform FE tasks with the assistance of humans and AI collectively. Our key contributions include:
\begin{itemize}
    \item We propose and design a human\&AI FE prototype, leveraging the benefits of both advanced AI- and human-generated knowledge to help DS practitioners with FE. 
    \item Taking the human\&AI FE as a research probe, we conduct a user study with 14 participants to understand their perceptions and experiences when working with the suggestions from humans and AI. 
    \item Based on our key findings, we provide practical design implications and research opportunities for developing user-friendly and effective human-AI collaborative interfaces/tools for DS practitioners to perform FE.
\end{itemize}

%% file: Sections/02-relatedwork.tex
\section{Related Work}  \label{related work}
Our work investigates how data science (DS) practitioners perceive and interact with a novel human\&AI FE design in real-world practice. 
We organize the literature into (1) data science project lifecycle and feature engineering, (2) automation for ML/DS (AutoML) and specifically for FE tasks (AI-assisted FE), and (3) interactive FE systems.

\subsection{Data Science Project Lifecycle and Feature Engineering} \label{human-relatedwork}
\begin{figure}[h]
  \centering
  \includegraphics[width=0.75\linewidth]{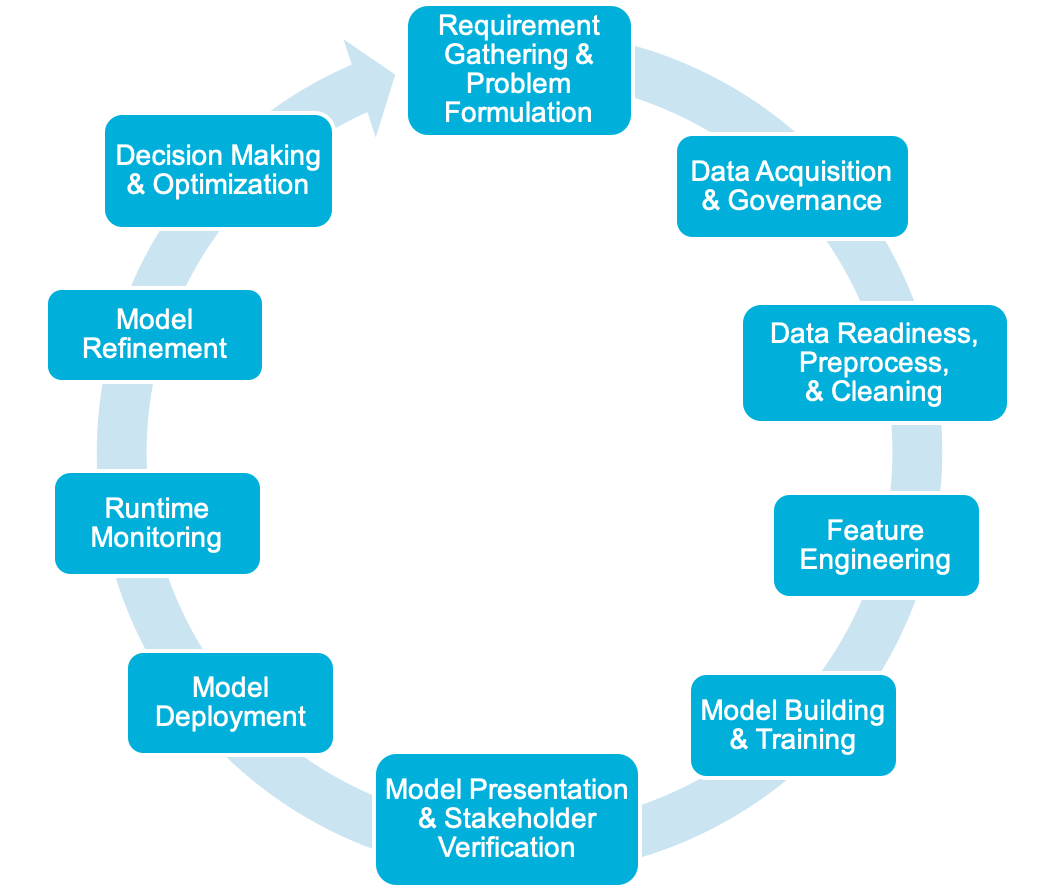}
  \caption{A 10-stage DS/ML lifecycle, starting at the top -- Requirement Gathering~\cite{wang2021autods}}
  \label{fig:AILifecycle}
\end{figure}

Data science is a multidisciplinary field that focuses on the processes for extracting insights from data~\cite{kross2019practitioners}. In data science practice, users often use machine learning (ML) techniques to build models that predict or provide recommendations based on input data~\cite{muller2019data}. Among the different stages in a DS project lifecycle shown in Fig.~\ref{fig:AILifecycle}, the most time-consuming stages are from requirement gathering to FE, which even accounts for 80\% of the entire project time~\cite{guo2011proactive, wang2021autods}. 
Numerous works have been conducted, and tools are being developed to support DS work in these stages before the actual model building begins~\cite{amershi2014power}. Examples of such efforts include data wrangling~\cite{guo2011proactive} and data augmentation~\cite{cashman2020cava}.

Feature engineering (FE) is one of the most crucial steps in the DS lifecycle ~\cite{aggarwal2019can, wang2021autods}. DS practitioners often need to spend considerable time creating features and experimenting with different combinations of these features to improve model performance~\cite{yu2021feature, muller2019data}. 
Given FE's interdisciplinary and collaborative nature, many HCI papers have focused on studying how to facilitate the collaboration among DS practitioners or between them and other roles (e.g., domain experts) in a DS project~\cite{piorkowski2021ai, Zhang2020collaborate, smith2020enabling, park2021facilitating, April2019Notebook}. 
Researchers have investigated the collaboration in DS teams and offered design implications for the collaborative context between DS practitioners using computational notebooks ~\cite{April2019Notebook}. 
A recent study analyzed the communication gap faced by AI developers and business owners, from which they suggested insights on how to promote communication and collaboration at various stages of the DS project lifecycle ~\cite{piorkowski2021ai}.

One notable recent work~\cite{smith2020enabling} proposed and evaluated a framework named \textit{Ballet}, which enables DS practitioners to collaborate in a crowdsourcing manner by committing their features and reusing others' features.
This approach facilitates the collaboration among multiple DS practitioners for FE when working on the same dataset. 
To facilitate communication and knowledge sharing between DS practitioners and domain experts, another work offered a design called \textit{Ziva}~\cite{park2021facilitating}. 
Although these works contributed promising perspectives and feasible systems for improving the efficiency of FE tasks, they all face a similar limitation --- addressing the ``cold start'' challenge. They require either an active open-source community or a large in-house DS team so that one user's FE task may reuse others' created features on a similar task. 
We have learned that many in-house DS projects may have only one or two DS practitioners~\cite{wang2019human,piorkowski2021ai}, and an open-source FE community may face various legal and IP issues. Thus, we still need to explore alternative solutions in addition to the human-human collaboration FE approach.

\subsection{Automated Machine Learning (AutoML) and AI-Assisted Feature Engineering} \label{AI-relatedwork}
AutoML refers to the group of algorithms and systems to automate the various stages of the DS lifecycle~\cite{zoller2019survey}. For example, researchers have proposed various works targeting the automation of the model building and training stage by searching for an optimal algorithm and combination of hyperparameter values~\cite{kotthoff2019auto}. 
Others focus on the automated extraction, transformation, and loading of datasets to support DS practitioners' data readiness and preprocessing tasks~\cite{kougka2018many}.

Specifically around the FE tasks in the DS lifecycle, and there has been a growing number of recent works that have developed various automated approaches for it ~\cite{kanter2015deep, khurana2016cognito, lam2017one}. 
These algorithms reported that they could generate a huge number of features within a very short amount of time.\footnote{Some algorithms can generate hundreds of features for a test dataset in a couple of minutes, whereas the same dataset may take DS practitioners a couple of days to manually create features.} 
For example, Kanter et al.~\cite{kanter2015deep} developed the deep feature synthesis (DFS) algorithm to generate features for relational datasets automatically. Lam et al.~\cite{lam2017one} developed OneButtonMachine, a system that can automatically join multiple database tables and apply transformation techniques to extract useful features.

As AI-assisted FE has increasingly matured together with other AutoML techniques, researchers and industry practitioners have begun integrating all these different automation components into an end-to-end AutoML solution. 
Incorporating user-centered design aspects and graphical user interfaces, researchers claim these systems represent instances of human-AI collaboration systems~\cite{wang2021autods,gil2019towards, xin2021whither}.
However, multiple user studies have reported that DS practitioners may not like or trust these AutoML systems, as they can not understand or interpret how they work~\cite{wang2019human,drozdal2020trust}.
How to balance humans' input and automation in such ML tools warrants further exploration~\cite{kandel2012enterprise, yang2018grounding,hohman2020understanding, lee2019human}.

\subsection{Interactive Feature Engineering Systems} 
Interactive feature engineering focuses on the under-explored but critical FE stage~\cite{sambasivan2021everyone}. 
It offers humans more control and insights when using an AI-assisted FE tool, thus it can balance the trade-off between automation and humans' controls~\cite{krause2014infuse, zhao2019featureexplorer}. 
One example work presented the feature generation process in a visual diagnosis tool using tree boosting methods and performed feature generation by testing the combinations of different mathematical operations as well as checking the feature similarity ~\cite{liu2017visual}. 
Other works designed visual analytics (VAST) systems to support automated feature exploration in classification problems and regression tasks, such as ExplainExplore ~\cite{collaris2020explainexplore} and HyperMoVal ~\cite{piringer2010hypermoval}. 
Various techniques have been offered to assist the feature selection task, 
like dimension reduction and clustering for high-dimensional data~\cite{yang2003interactive, yang2004value}, correlation matrix ~\cite{rojo2020gacovi}, and ranking design ~\cite{zhao2018iforest, krause2014infuse}. 
These works employed different metrics to measure the linear and nonlinear relationships between generated features, including the Pearson correlation and mutual information ~\cite{rojo2020gacovi, may2011guiding, chatzimparmpas2021featureenvi}.

However, all these aforementioned works were provided as independent visual analytical systems that separate from existing DS work environments (e.g., online notebooks that millions of DS practitioners use in their daily work~\cite{kluyver2016jupyter,rule2018exploration, kery2018story}). 
In addition, they did not consider the novel and complex dynamics and user patterns enabled by the combination of human-assisted and AI-assisted feature engineering approaches.
In this work, in addition to presenting the human\&AI FE prototype, we also explore how DS practitioners perceive and use the suggested features, which are presented as a notebook plugin in real-world practice.


%% file: Sections/03-framework.tex
\begin{figure*}[htbp]
  \centering
  \includegraphics[width=\linewidth]
  {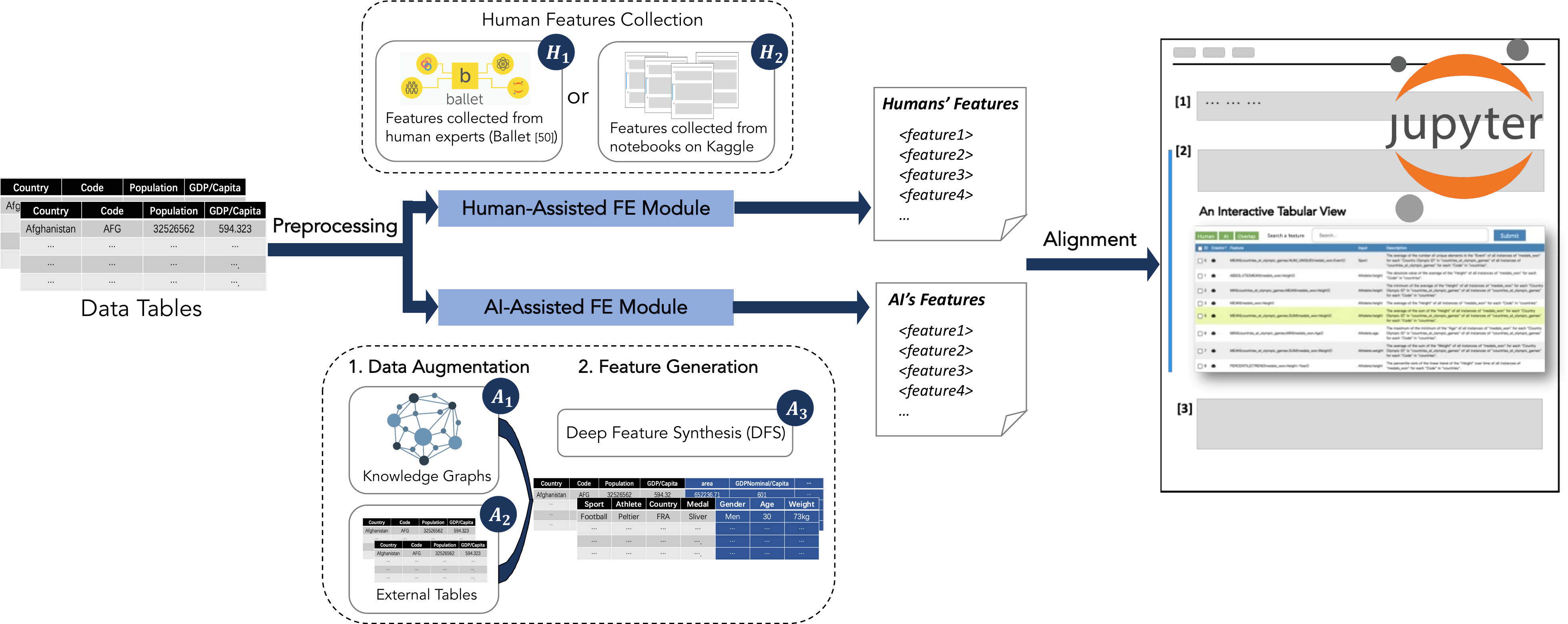}
  \caption{An overview of the human\&AI FE design that synthesizes human- and AI-generated features collectively in Jupyter notebooks. The whole pipeline runs from left to right. 
  Taking the given data table as input, we first process it (i.e., data cleaning), and then input the table into the human-assisted module and AI-assisted module separately. Each module generates a list of features, which can be organized in online notebooks using a plugin and displayed with an interactive tabular view for feature recommendation. 
  } 
  \label{fig:framework}
\end{figure*}
\section{Human\&AI Feature Engineering Design}  \label{framework}
To explore how DS practitioners use and perceive the human\&AI FE, we implement a usable prototype to integrate feature suggestions based on the state-of-the-art feature engineering (FE) practices of humans~\cite{smith2020enabling} and AI ~\cite{kanter2015deep, khurana2021semantic}. 
It serves as a research probe that aims to understand users' perception and behavior of the human\&AI FE paradigm. 
As shown in Fig.~\ref{fig:framework}, the human\&AI FE takes data tables as the input and synthesizes human- and AI-suggested features through two separate modules. It then aligns the two sets of resulting features and presents them to users in the form of a notebook plugin -- an interactive interface embedded in the code blocks -- in the Jupyter Lab. Jupyter Lab has been widely used by the data science community ~\cite{perkel2018jupyter} and can guarantee easy and broad adoption ~\cite{granger2016jupyterlab}.

Before going into the details, we first take an example to illustrate the feature recommendation from humans and AI. Assuming that we have a dataset containing $user\_id$, $body\_height$, and $body\_weight$ three columns. 
AI-assisted FE may generate features such as the square root of body height ($sqrt(body\_height)$) or the difference between an individual's body weight and the group average ($body\_weight-mean(body\_weight)$).
Humans can generate features with the specific knowledge, such as body mass index ($BMI=body\_weight_{lb} / {body\_height_{in}}^2 * 703$). 
Next, we elaborate on the design rationales of each component in the human\&AI FE in the following sub-sections.

\subsection{AI-assisted Feature Engineering Module}  \label{automated}
We incorporate two methods to achieve automated feature engineering: 1) automated data augmentation with external knowledge (Fig.~\ref{fig:framework} (A1)(A2)), and 2) deep-feature synthesis (DFS) ~\cite{kanter2015deep} to automatically generate large-scale features (Fig.~\ref{fig:framework} (A3)). 
We use external Knowledge Graph (KG) and tables to enrich the input data tables. It is a common augmentation method used in machine learning (ML)~\cite{hulsebos2021gittables, cashman2020cava} as enhancing input data with externally related information could help the AI generate diverse features~\cite{galhotra2019automated, lam2017one}. 
We then apply the Deep Feature Synthesis (DFS) algorithm to generate large-scale features, which has been used for mature feature engineering platforms (e.g. the Featuretools~\footnote{\label{ft}https://www.featuretools.com/}) and has been proved to have stable performance with real-world datasets~\cite{kanter2015deep}.
\begin{figure*}[htbp]
  \centering
  \includegraphics[width=\linewidth]
  {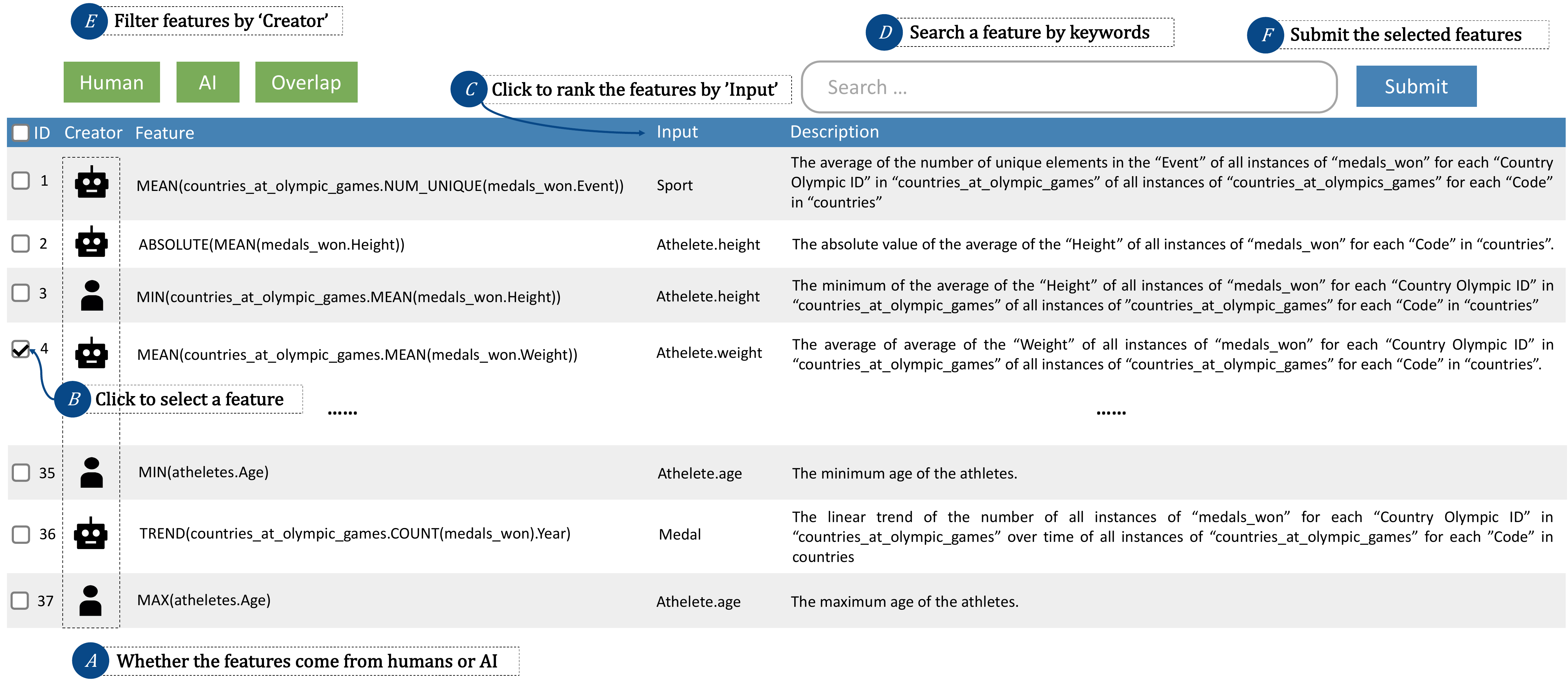}
  \caption{The interactive tabular interface embedded in the plugin of Jupyter Notebook. Users could interact with the view to check, filter, and select features suggested by other humans and/or AI.} 
  \label{fig:interface}
\end{figure*}

\subsubsection{Automated Data Augmentation with External Knowledge} \label{dataAug}
As shown in (A1) and (A2) in Fig.~\ref{fig:framework}, we leverage two kinds of external knowledge to augment the input data. To augment the table with new data attributes in KGs, we use Table2KG~\cite{nguyen2020mtab4wikidata, Ernesto2020semanticWeb} to map attributes in the input data table(s) to semantic entities in external knowledge graphs (KGs) (e.g., DBpedia\footnote{DBpedia (``DB'' for "database") offers knowledge graphs that allow users to semantically explore the Wikipedia sources by querying relationships and properties} used in this paper). 
After that, we extract all properties of the associated KG entities and add the new ones as attribute columns into the original data tables. Then, for each data item, we automatically fill its value, of the new attribute columns by querying the KGs. Finally, we keep the newly added attributes whose ratio of null values is smaller than 40\% to reduce the impact of the data sparsity problem of KGs ~\cite{ji2021survey}.
In addition, we exploit related data tables on GitHub or Kaggle as another external source for augmentation by searching the matched attributes in table headers ~\cite{hulsebos2021gittables} (Fig.~\ref{fig:framework} (A1)).

The two methods (i.e., using KGs and external tables) for data augmentation can be automatically performed with the preset parameters. In reality, users can also replace these resources with their data repositories and/or other general or domain-specific knowledge data.

\subsubsection{Automated Feature Generation with Deep Feature Synthesis}
Taking the augmented data tables as input, we use the Deep Feature Synthesis (DFS) algorithm~\cite{kanter2015deep} to generate new features automatically (Fig.~\ref{fig:framework} (A3)). 
DFS can create new features by (1) leveraging the relational attributes (e.g., \textit{country} and \textit{GDP}) among multiple data tables and (2) applying various transformations of features based on these attributes and their data types (e.g., numeric or continuous).  
We take the result of DFS as the output of AI-suggested features.

\subsection{Human-Assisted Feature Engineering Module} \label{humanFeature}
\subsubsection{Two Methods for Collecting Human Created Features}
We use two methods to collect human-generated features, as shown in Fig.~\ref{fig:framework} (H1) and (H2). 
Both methods are practical for gathering features from DS practitioners~\cite{smith2020enabling, muller2019data} and they can be complementary according to the number of available contributors. 
One is using Ballet ~\cite{smith2020enabling} -- an online collaborative FE design -- that supports multiple DS practitioners to submit and reuse features online based on the same dataset. 
The other is manually collecting human-generated features by experienced DS practitioners from open-source platforms, such as Kaggle and Github (Fig.~\ref{fig:framework} (H2)). 
These two methods simulate two paradigms of how FE can be jointly performed by humans in reality. 
Ballet exemplifies a \textit{``feature crowdsourcing platform''}~\cite{smith2020enabling} in which a large number of experienced DS practitioners contribute their handcraft features and learn from others' proposed features. 
The second approach demonstrates the collaborative efforts of compiling high-quality, hand-crafted features by several DS practitioners (usually two or three) in a data science team. 
In practice, users could choose between these two methods based on the available source of DS practitioners.

\subsubsection{Feature Format and Information Collection} \label{Feaformat}
We followed the feature format used in Ballet ~\cite{smith2020enabling} to align human- and AI-generated features.
This format contains three kinds of information: (1) the definition, (2) the input attribute, and (3) the textual description. The information has been used as a standard to collect human features \cite{smith2020enabling}.
The definition of a feature is determined by its \textbf{transformation formula}, which expresses how a feature is created from the input data attribute(s) in the original data tables. 
The input attribute of a feature is the input data columns in the augmented data table (Sec.~\ref{dataAug}). The description of a feature is a short text that explains its semantic meaning. The information on human-generated features can be easily collected using some collaborative tools (e.g., Ballet) or analyzing notebooks by experienced DS practitioners. For AI-generated features, we extract the three types of information from DFS results automatically by open source tools\footnote{\url{https://featuretools.alteryx.com/en/stable/api_reference.html}}.

Taking the Olympic Games Medal dataset as an example, the \textbf{input attribute} ``medal\_won'' represents the number of medals won by each country in a historical Olympic Game. A new feature may first count the total number of medals over a period using the \textit{``SUM()''} transformer, and then, calculate the mean value of medals using the \textit{``MEAN()''} transformer. Hence, the \textbf{definition} of this new feature is ``\textit{MEAN(countries\_at\_olympic\_games.SUM(medals\_won))}''. The \textbf{description} of the feature is \textit{The average medals won by each country over the period.}

\subsection{Feature Recommendation User Interface} \label{extension}
We implemented a simple and practical feature recommendation interface in an interactive tabular view as DS practitioners are familiar with using tables to view data in the online notebooks~\cite{siyuan2021KTabulator, cashman2020cava}. 
The information is automatically processed and loaded from the backend and then is shown in the tabular view, based on the feature format in Sec.~\ref{Feaformat}.
As shown in Fig.~\ref{fig:interface}, the view presents five types of information for each feature, including (1) feature ID, (2) creator, (3) feature definition, (4) input attribute, and (5) the description of a feature. 
In order to fit the human\&AI-assisted FE into real-world practice for DS practitioners, we implement the interface as a plugin of Jupyter Notebook for users to check the information of features and directly select features to test their performance in the following ML models. The interface of the plugin can be triggered by clicking a button embedded in Jupyter. We believe this kind of design can offer great generalizability without changing the existing DS pipeline.

Users can select a feature or all features by the checkbox next to \textit{feature ID} (Fig.~\ref{fig:interface} (B)). The column next to the feature ID shows the ``Creator'' of a feature (Fig.~\ref{fig:interface} (A)) by icons, like \AutoIcon\ or \humanIcon, indicating the feature was created by AI or a human. 
With this information, users can filter features by clicking the green buttons (Fig.~\ref{fig:interface} (E)) to access or select all humans' or AI's features separately. 
In addition, users can click the text in the header to rank features in either ascending or descending alphabetical order (Fig.~\ref{fig:interface} (C)). We also provided a search box for users to search features by inputting the keywords of feature definition (Fig.~\ref{fig:interface} (D)). 
After selecting features, users can click the \textit{``submit''} button to generate an array of selected feature(s) automatically in the notebooks (Fig.~\ref{fig:interface} (F)) and the features will be processed as the input for the following ML model(s).

%% file: Sections/04-study.tex
\section{User Study} 
We conducted a case study with 14 data science (DS) practitioners who have project experience in data science and feature engineering. 
This study goal is to conduct an exploratory study to understand investigate how DS practitioners integrate the proposed human\&AI-assisted FE into their work practices (\textbf{RQ1}) and explore their needs (\textbf{RQ2}) and the potential challenges (\textbf{RQ3}) in incorporating human\&AI FE into data science lifecycles.
During the study, we allowed the participants to use features generated from both humans and AI by taking them as potential collaborators, and they can also create new features by themselves.

\input{table/DSworkers}
\subsection{Participants}
With institutional IRB approval, we recruited participants through snowball sampling by posting recruitment messages on mailing lists, social media, and personal connections. 
After seven days of recruitment, in total, 14 participants (4 females) were between the ages of 24 to 39 and had experience in machine learning and computational notebooks (e.g., Jupyter Notebook or Google Colab).  
As shown in Table ~\ref{tab:participant}, the participants had on average seven years of programming experience and four years of experience in data science. Six of them are citizen practitioners without career experience (doctoral candidates major in AI), and the others are from the industry. 
The \textit{Project Exp.} column shows their recent project experiences across different application domains, including healthcare, finance, etc. 
All participants have never used features from both humans and AI collectively before.
Each participant received \$20 as a reward for participating in the study.

\begin{figure*}[htbp]
  \centering
  \includegraphics[width=\linewidth]{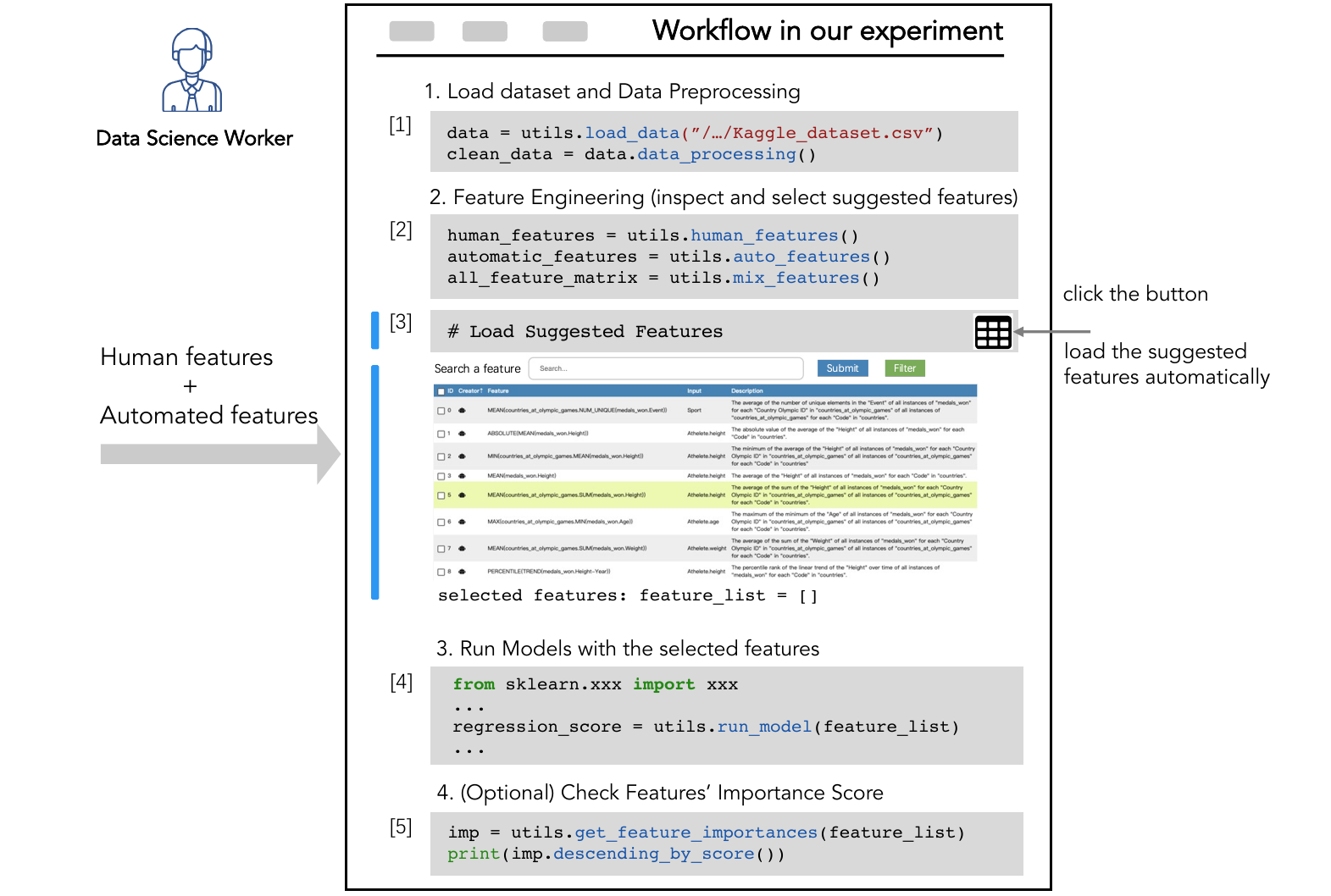}
  \caption{The experimental notebook interface with certain helper code snippets (e.g., data loader and model training) that each participant used in our study. They followed the workflow and experienced the human\&AI FE to choose, analyze or compare the suitable feature set for the given task.} 
  \label{fig:study1}
\end{figure*}

\subsection{Feature Preparation for the Study} \label{dataPreparation}
To verify the reliability and stability of the proposed human\&AI FE prototype, we tested various datasets\footnote{\label{house}Predict house prices: \url{https://www.kaggle.com/c/house-prices-advanced-regression-techniques}}\footnote{\label{census}Predict census income: \url{https://mit-dai-ballet.s3.amazonaws.com/census/ACS2018_PUMS_README.pdf}}\footnote{\label{outcome}Predict life outcomes: \url{https://opr.princeton.edu/archive/FF/}}\footnote{Predict Olympic medal:\url{https://www.kaggle.com/heesoo37/120-years-of-olympic-history-athletes-and-results}\label{olympic}} as the input in pilot sessions. 
We transformed these data tables to \textit{dataframes} using \textit{Panda}\footnote{\label{panda}https://pandas.pydata.org/} before constructing features using the human\&AI FE. 
For the in-depth qualitative investigation, we chose one of the tested datasets and the corresponding competition on Kaggle\footref{olympic}, which is about the modern Olympic Games with the task of predicting the medals won by each country. The Olympic game topic is familiar to the public and it is a well-known international event. 
In addition, we found that there are few features proposed by DS practitioners under this competition, leaving more space for us to explore new features from both humans and AI.


We collected human features based on the second method in the proposed human\&AI FE pipeline (Fig.~\ref{fig:framework} (H2)). The method is suggested by two data scientists working with us and it is a common way used by DS practitioners to manually collect features online~\cite{wang2020autoai}. Two co-authors with sufficient experience in Kaggle competitions ($MEAN = 3.5$ years) participated in the feature collection process. 
We first collected the online notebooks under the competition\footref{olympic} and ranked them by the number of votes. We then selected the top 5\% notebooks to collect human-generated features based on the data scientists' suggestion. 
Although some users share the code of data processing and constructing features on Kaggle, we found that the human-generated features online are still rare.
We selected all the human-generated features with sufficient justifications and detailed explanations of their creation process and examined their code to ensure the quality and credibility of features, and finally obtained 10 as the human-assisted features (Sec.~\ref{humanFeature}). 

For AI features, we used the data tables about \texttt{TABLE 1: countries} and {\texttt{TABLE 2: winning records of athletes}} from the selected dataset. 
Since AI could automatically generate a large scale of features in a short time, we expected these AI-generated features to compensate for the scarcity and simplicity of human features. 
Following the augmentation process described in Sec.~\ref{automated} (Fig.~\ref{fig:framework} (A1) and (A2)), we added 23 new data columns from the external knowledge graphs and tables and inserted them into the input data tables. 
These new data attributes contain information related to the economy and population of a country, such as the total population or the GDP per capita, and the athletes' personal information (e.g., age).
In total, we got three tables, 32 data attributes, and 134732 data instances (as the rows).
Finally, we applied the DFS algorithm to the augmented tables to generate new features (Fig.~\ref{fig:framework} (A3)). The parameters of the DFS algorithm were set according to the API documents\footnote{\url{https://featuretools.alteryx.com/en/stable/generated/featuretools.dfs.html}}. We added the information for each feature following Sec.~\ref{Feaformat} and presented them in the interactive tabular view as a plugin view of Jupyter notebooks (Sec.~\ref{extension}).

In summary, we obtained 99 features from the human\&AI FE, comprising of 10 human-created features and 89 AI-generated features. The number of features collected from both humans and AI was generated in a practical situation without deliberate controls, which signifies that AI could produce more features than humans with the same input. 
This also aligns with the real-world observation that it is generally more challenging to acquire human features than AI features from the DS practitioners' perspective~\cite{sambasivan2021everyone, muller2019data}. These collected features are presented in a random order in the table during the study.

\subsection{Experiment Environment: Notebook and Server Setup} \label{config}
\subsubsection{Workflow} \label{workflow}
To help participants focus on the FE task and spend less time on other parts of the data science workflow, we constructed a Python-based notebook (Fig.~\ref{fig:study1}).
The experiment notebook is a solution to this particular Olympic dataset that we collected and aggregated from multiple winning solutions on Kaggle\footnote{\url{https://github.com/Featuretools/predict-olympic-medals}}. 
The entire code is executable and can deliver a model result even without any further coding. 
The model performance result will be updated once participants change the feature selection in the FE section.
We divided the notebook code blocks into four parts: 
\begin{itemize}
    \item \textbf{Loading and preprocessing data}: We pre-loaded the data tables after data cleaning for participants to check the input data attributes. 
    \item \textbf{Loading and presenting the suggested features with a tabular view}: We offered codes for loading features generated by humans and AI based on the human\&AI FE (Sec.~\ref{framework}). Participants can check the feature construction process or access the meta-data of features in \textit{DataFrames} storing the values of each feature in a matrix. They can inspect the information on the suggested features and select features through a tabular view. 
    \item \textbf{Running the model}: Then, participants could execute the following code blocks to check the performance of selected features with a random forest model~\cite{pal2005random}, which has been suggested and tested by the data science expert working with us.
    \item \textbf{Checking the performance scores of selected features}: We also provided a code block that provides a model performance evaluation function. It can generate the performance scores of ML models and display the contribution value of each feature. 
\end{itemize}

\subsubsection{Task and Notebook Setup}
The task for participants involved is to find the best subset of features and give their reasons why they choose the feature(s) with the given medal prediction task and dataset. 
We mainly focus on how DS practitioners use and adopt features suggested from other human collaborators and AI and we also allow them to create features using our system.
We did not strictly constrain how each participant should define the \emph{best} subset of features because we aimed to investigate how they use our proposed human\&AI FE naturally in a real-world scenario.

We published the notebook installed with the plugin of our proposed human\&AI FE online and deployed it on a cloud server so that participants could access the environment simply through a shared link in the study without configuring any coding environment.

\subsection{Experiment Procedure}
The study was conducted remotely via Zoom due to pandemic restrictions in the past. It comprised an experimental session and an interview, averaging around 69 minutes in total.

\subsubsection{Experiment Session}
During the experiment, we initially introduced the task to participants. Once consent was obtained, we gathered demographic data and information about their prior experience in FE. 
Participants were then given a unique link to access the experiment notebook server and user study interface (Sec.~\ref{workflow}). After opening the notebook, we requested permission to share their screen and began recording the session. We then guided them through the interface to familiarize them with the overall workflow and the functions provided in the human\&AI FE interface.
We did not disclose the detailed feature collection process to participants to avoid cognitive biases.

During the study, we encouraged participants to optimize their model performance. They were free to test numerous feature combinations, and upon completion, they submitted their final feature selection and associated model performance. Participants were allowed to explore, modify the code, ask questions related to the experiment, and were encouraged to verbalize their thoughts. They were free to conclude the session whenever they were satisfied with the ML model performance and feature selection. On average, the experiment sessions lasted for around 46 minutes.

\subsubsection{Post-experiment interview and questionnaire} \label{interview} 
Upon completion of the main study, we carried out a semi-structured interview (averaging 23 minutes) with each participant to gather their usage patterns, perceptions, and suggestions for the human\&AI FE. The interview primarily included three open-ended questions (Q1-Q3). Based on participants' responses, we asked tailored follow-up questions for each participant, considering their responses and our observations of their behaviors.
\begin{itemize}
    \item \textbf{(Q1)} What are your strategies for selecting features during the study, and why do you select these features?
    \item \textbf{(Q2)} How do you like the features suggested by humans and/or AI, both about the overall workflow and the user interface?
    \item \textbf{(Q3)} Based on your experience, could you give us some suggestions on how to further improve the usability and user experience of the human\&AI FE? 
\end{itemize}
Beyond the above questions, we also invited each participant to share their opinions on the challenges of their daily FE tasks to further explore their requirements.


\subsection{Data Collection and Analyses} \label{data-analysis}
\subsubsection{Feature Selection}
To investigate how participants select features from both humans and AI (RQ1), we tracked each participant's feature selection process and their notebook editing history. Particularly, we recorded the final feature selection of all participants, encompassing the feature ID(s), feature definition(s), the creator(s), input attribute(s) of the selected feature(s), and the corresponding model performance.

\subsubsection{Post-questionnaires}
We designed a post-experiment questionnaire to collect participants' perceptions of both human- and AI-generated features across four dimensions: trust, perceived explainability, and perceived performance of these features. 
We collected feedback of the four dimensions by asking questions using AI and human as the creator of features separately.
To analyze the participants' responses, we utilized a nonparametric test (Mann-Whitney U)~\cite{mcknight2010mann} as the data do not follow a normal distribution, and reported the results in Sec.~\ref{RQ2-difference}, alongside participants' qualitative feedback. 
Additionally, we gathered participants' views on the overall human\&AI FE, its interactive interface, and the perceived task difficulty. All the questions in the post-questionnaire utilized a 7-point Likert scale (1: strongly disagree, 7: strongly agree).

\subsubsection{Interview Logs}
We analyzed interview responses to identify participants' feature selection strategies (RQ1), their experiences and perceptions (RQ2), as well as their suggestions for improving the human\&AI FE (RQ3).

\subsubsection{Analysis Methods}
Three authors performed open coding~\cite{charmaz2006constructing} on the audio transcripts of all participants. Initially, we independently analyzed the data to generate preliminary codes. Subsequently, we held three rounds of discussions to compare, group, and refine these codes, resulting in a final code book.
Two authors carefully reviewed the code themes to consolidate them into categories and eliminate any irrelevant ones. After multiple rounds of discussion, we reached consensus and organized the findings in line with our research questions. 

%% file: table/DSworkers.tex
\renewcommand{\arraystretch}{1.5}
\begin{table*}[tp]  
  \centering  
  \fontsize{8}{8}\selectfont

\caption{Participant demographics. The first column shows the participant ID (PID). The second column displays participant's experience in programming, machine learning and their working experience in industry (years). Participants’ roles and their shared project experience is shown in the fourth and fifth columns.} 
 \label{tab:participant}
\begin{tabular}{p{0.5cm}<{\centering}p{2.5cm}<{\centering}p{3cm}<{\centering}p{5cm}<{\centering}p{0.5cm}<{\centering}}

\toprule
PID&Years of Exp. (Coding/DS/Job)&Role&Project Exp.\\ \hline 
P1&9/6/1&Citizen Data Worker&Aesthetic Evaluation of Images\\ 
P2&8/5/2&AI Engineer&Text Classification for Online Messages\\ 
P3&5/5/2&AI Engineer&Prediction of E-commerce Transactions\\ 
P4&5/3/1&Citizen Data Worker&Prediction of House Price\\ 
P5&6/3/1&Citizen Data Worker&Behavior Detection in Games\\ 
P6&8/6/5&Expert Data Scientist&CPU Memory Prediction\\ 
P7&9/2/1&Citizen Data Worker&Object Detection in Images\\ 
P8&11/5/1&Citizen Data Worker&Question Answering in Paragraph\\ 
P9&11/7/7&Expert Data Scientist&Weather Storm Prediction\\ 
P10&8/4/1&Citizen Data Worker&Question Answering System\\ 
P11&9/6/3&ML Engineer&User identification on Social Media\\ 
P12&8/4/4&Expert Data Scientist&Financial Credit Scoring\\ 
P13&6/6/6&ML Engineer&Medical Image Classification\\ 
P14&11/6/5&ML Engineer&Prediction of Medical Data\\ 
\bottomrule
\end{tabular}
\label{tab:participant}
\end{table*}

%% file: Sections/05-result.tex
\section{Results}
We begin with an analysis of participants' feature selection, followed by their perceptions and experiences with the joint recommendation of human and AI-generated features. Lastly, we identify user requirements to improve the usability of human\&AI FE.

\subsection{\textbf{Feature Selection Result (RQ1)}} 
\begin{figure*}[htbp]
\centering
  \includegraphics[width=0.9\linewidth]{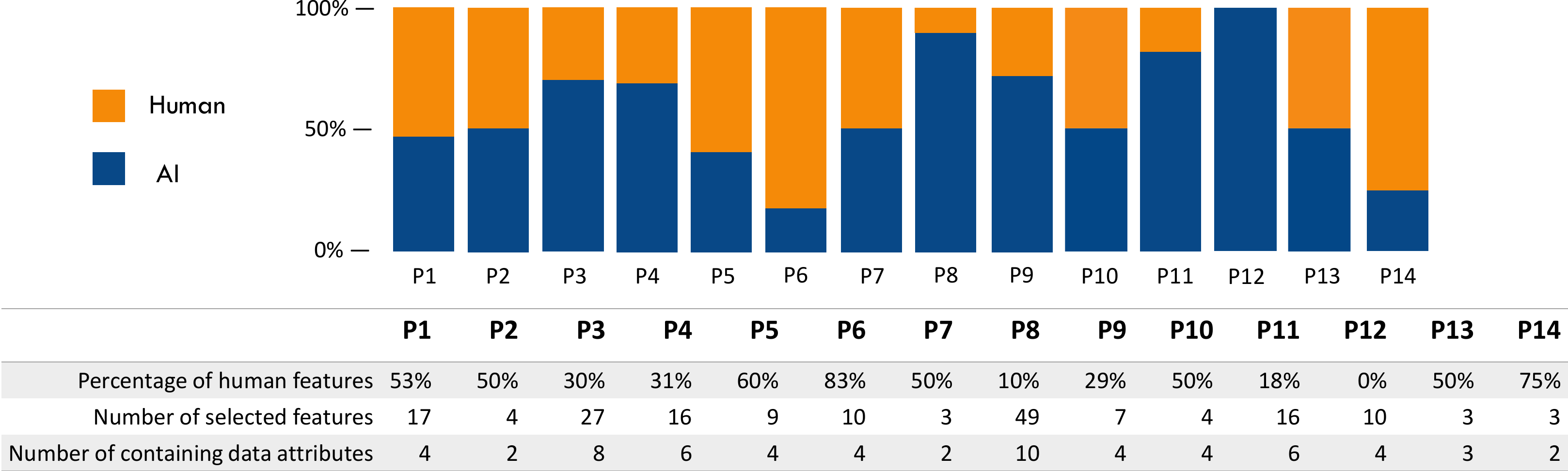}
  \caption{Participants' final selection of features generated from humans and AI. The stacked chart shows the percentage of features selected by each participant. The table presents details of the selection results, including the number of selected features, the number of containing data attributes in the input data tables, and the percentage of human-generated features.}
  \label{fig:result3}
\end{figure*}

Fig.~\ref{fig:result3} presents participants' final feature selections.
The details for each participant include (1) the percentage of features from humans and AI in a stacked bar chart, (2) the number of selected features, and (3) the number of data attributes from the original data tables covered by the selected features. 
Participants were asked to submit results they deemed satisfactory.
On average, the participants performed about eight iterations to reach their desired results, with the output in a relatively small range: from 0.780 to 0.871 ($MEAN = 0.821, SD = 0.022$). 
Most participants ceased further attempts upon achieving around 80\% performance, while a few (P2, P5) exceeded 10 iterations and found little variation in the output performance.

From the selection outcomes, we observed that a majority of participants (13 out of 14) \textbf{chose features suggested by both humans and AI} for their final selection. Only one participant (P12) exclusively selected AI-generated features, citing that they provided significant inspiration, resulting in all their time being spent exploring AI-suggested features. 
Additionally, we found the number of covered data attributes in the selected features may related to the percentage of selected AI features with a marginally significant trend ($r = 0.524, p-value = 0.054, Pearson's\ R$), considering that AI-generated features may contain more diverse data attributes than those from humans.
Generally, participants' choice of features can be affected by multiple factors, and we analyze the strategies in the following subsections.

\subsection{\textbf{Feature Selection Strategy (RQ1)}} \label{strategy-creator}
\subsubsection{Selection Strategies based on Feature \textit{``Creator''}.} \label{RQ1-2}
In our design, we included a \textit{``Creator''} column in the interactive tabular view (Sec.~\ref{extension}), signifying whether a feature originated from AI or a human.
Our objective was to examine whether the \textit{``Creator''} attribute influences participant selection and how this information is used in their feature choice process (responses to Q1 in Sec.~\ref{interview}). We gathered the results from participants' feedback and corroborated their reflections with their actual behaviors by reviewing the video recordings of the experiment.

During the study, we identified \textbf{three distinct strategies that participants used in relation to the \textit{``Creator''}}: (1) selecting features strictly based on the \textit{``Creator''}, (2) using the \textit{``Creator''} as a reference in feature selection, and (3) disregarding the \textit{``Creator''}. 
These behaviors reflect varying degrees of reliance on the feature source in their decision-making process. Four participants (P1, P5, P6, P11) considered the \textit{``Creator''} of features critical, basing their final selection on it, as they placed more trust in features generated by humans. For instance, P6 said that \textit{``I am of the belief that these experts have likely conducted in-depth domain research and leveraged their knowledge to create these features.''}
Similar to P6, other participants considered it essential to identify whether a feature was derived from human experts. This distinction allowed them to grasp the underlying rationale of feature creation by adopting the perspective of these experts. This finding implies that certain DS practitioners might lean towards using human-generated features, valuing peer knowledge over AI-generated features.

Secondly, half of the participants (P2, P3, P4, P7, P8, P9, P10) reported that they mainly used the \textit{``Creator''} as a reference for them to check and test features: \textit{`The `Creator' did influence my selection process, but not the final outcome. I chose to test features separately from both humans and AI to assess their individual contributions to the results. (P3)'' }
Thus, the ``Creator'' label serves as a tool that aids certain participants in comparing and evaluating various feature combinations during their work, rather than directly influencing their final decisions. As half of our participants indicated a preference for considering the ``Creator'' as a reference during the task, this presents new opportunities for future design considerations in the realm of human\&AI FE (refer to Sec.~\ref{dis-6.1} for further details).

For the three participants (P12, P13, P14) who said that they did not care about the \textit{``Creator''} when choosing features with the human\&AI FE, they took all features as the same. 
P14 mentioned that \textit{``What matters is to know how the recommended features were generated. I do not discriminate between AI features and human features if they are easy to understand.''}
They mostly used feature definition or other semantic information to make decisions.
This kind of feedback from our participants raises critical problems about the explainability of features from both humans and AI. We also found participants' perceived differences in the explainability of human and AI features and present the result in Sec.~\ref{RQ2-difference}.

In conclusion, we find that participants treated the ``Creator'' of features differently and whether a feature comes from AI or humans can play a role in most users' FE process. Future design can consider whether and how to provide the ``Creator'' information based on users' requirements to support their inspection in collaborative human\&AI FE.

\subsubsection{Selection Strategies based on Perceived Semantic Meanings of Features.} \label{RQ1-3}
During the study, all participants considered the definition and description of features when inspecting them, basing their selection primarily on the semantic meaning of each feature. 
However, participants \textbf{perceived and utilized the semantic meaning of features from different perspectives}. For example, participants P2, P3, and P4 preferred to select features that aligned with their background knowledge. According to participant P2 \textit{``I prefer to choose the features with $MAX()$ transformer for the \textit{Athlete.height} and \textit{Athlete.weight}, and use $MIN()$ when the feature has \textit{Athlete.age}.''}
In addition to considering concrete elements in a feature, there are also some other participants (P5, P6, P11) who cared more about the explanation of a suggested feature as they needed to \textbf{understand how and why a feature was created}. For example, as reported by participant P5 \textit{``I make an effort to comprehend the construction process and underlying motivations behind the features created by human experts.''} 

Several participants (P1, P9, P11, P12, P14) further gave the reason why they prefer to choose explainable features. First, they may need to \textbf{explain the semantic meaning of features to the stakeholders}, such as their customers or collaborators, in work practices. Second, explainable features are perceived to be \textbf{more reliable and trustworthy}. Participant P1 highlighted that \textit{ ``I select features that I can interpret or, at the very least, understand, as it is my responsibility to justify the model and feature selection to product managers or other stakeholders, in order to convince them to adopt my solution.''}


In summary, the semantic meanings and explanations play a crucial role in DS practitioners' feature selection process. It is evident that they do not solely rely on performance when choosing features. Rather, they consider the interplay between explainability and performance to make informed decisions. 
For instance, DS practitioners expressed difficulties in comprehending certain high-performance features generated by AI, ultimately leading them to discard such features. Therefore, if our objective is to encourage users to embrace recommended features, merely emphasizing their performance or contribution is insufficient to persuade users. Instead, meaningful explanations or descriptions of the features should be provided to enhance their understanding and acceptance.

\subsubsection{Different Strategies Between Citizen Practitioners and Professional Data Scientists.} \label{RQ1-4}
In addition to the above strategies, we obtained an interesting finding that \textbf{participants' working experience may affect their feature selection process as well as the final decisions}. 
However, this discrepancy in background knowledge was not reflected in other aspects according to the findings. 
Thus, we analyzed the background of each participant and divided them into two groups. Participants who have more than 2-year of work experience or those whose roles are data scientists are arranged in the data scientists group, and the others were assigned to the citizen practitioners group \textit{(professional:citizen = 8:6)}.

We identified three possibile differences in feature selection and feedback between the two groups. Firstly, participants in the professional group indicated a trend of selecting a rather lower number of features ($M = 9.63, SD = 8.40$) compared to those in the citizen group ($M = 16.33, SD = 17.04$). These professional data scientists reported that they considered the representativeness of features to avoid redundancy. \textit{```I examined the feature information and conducted performance tests, which revealed that only a small number of features made a significant contribution to the model accuracy (P9).''}
The second observed difference is that data scientists (3/8) may tend to \textbf{try various combinations of features}, while participants in the citizen group may \textbf{test features with similar semantic meanings}. One potential reason is that the less experienced participants may lack the awareness of including diverse features, which is consistent with the point above. As participant P7 in the citizen group pointed out, \textit{``I found the features related to athletes' age could improve the performance, so I added more features about the age attribute.''}
The third observed difference is that some data scientists (P9, P11, P12, P14) tend to took the AI-generated features \textbf{collectively}, while participants (P4, P5, P7, P8, P10) with less experience tended to view the AI-suggested features \textbf{individually}. 
For example, participant P11 in the professional group pointed out, \textit{``I always take AI features as a whole because I think it is hard to explain a single AI feature. They are not intuitive  with a complicated transformation.''}
While participant P10 in the citizen group said that \textit{``I tried replacing the AI feature by another one to test which one could work.''} Other data scientists reported that they would treat AI- and human-generated features relatively equally.

These findings highlight the importance of tailoring feature recommendation support based on users' experience in feature engineering. For example, the human\&AI FE system should take into account users' background knowledge and their decision-making behaviors during the feature engineering process. In cases where users lack sufficient experience in feature engineering, the system should provide appropriate guidance and support, such as encouraging the inclusion of diverse features when users tend to select only similar features recommended by the system.
Moreover, future research needs to to identify commonalities among users while also offering a flexible user interface that can accommodate individual differences (see Sec.~\ref{dis-6.4}).

\subsection{\textbf{Perception of Using the Human\&AI FE(RQ2)}} 
\subsubsection{Perceived Difference of Features from Human and AI}\label{RQ2-difference}
In this and the following sections related to RQ2, we report the participants' perception and their experiences of human\&AI FE (RQ2) in three aspects, including (1) the perceived difference and (2) perceived complementarity between human- and AI-generated features, (3) and their feedback on the way we suggested and presented features (our interface design), as well as the overall workflow of human\&AI FE.
\begin{figure}
\centering
  \includegraphics[width=\columnwidth]{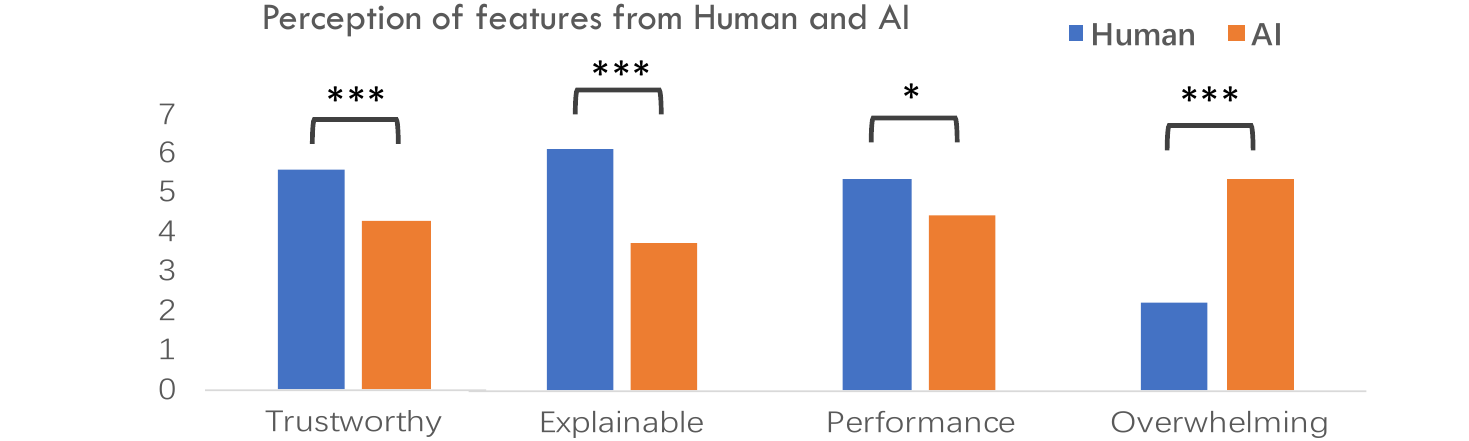}
  \caption{Results of the post-questionnaire about participants' perceptions of features from humans and AI.}
  \label{fig:result2}
\end{figure}

As shown in Fig.~\ref{fig:result2}, one prominent distinction between features from humans and AI is the perceived explainability, which aligns with the findings from the post-questionnaire analysis ($t(26) = 5.89, p < .01$). The majority of participants (10 out of 14) expressed difficulties in comprehending the explanations provided for AI-suggested features, particularly in terms of the definitions and descriptions. Participants expressed a preference for "human-like" descriptions that are more intuitive and informative, rather than relying on generic or ``templated'' explanations. 

In addition, some participants (P1, P4, P5, P7, P10, P11) mentioned \textbf{the complexity of the AI-generated transform functions}. For example, P4 complained that (s)he could not figure out the functions defined by AI and interpret them clearly.
The level of explanation and complexity of the suggested features significantly influences people's trust in them. The majority of participants exhibited higher trust in human-suggested features compared to AI-suggested features, as evidenced by the results of perceived trustworthy ($t(26) = 3.61, p-value < .01$). 
This finding is consistent with previous research ~\cite{arnold2019factsheets, drozdal2020trust, xin2021whither}. 
Moreover, it had a noticeable impact on the feature selection process and their perceived performance of the features from humans and AI ($t(26) = 2.218, p-value < .05$). In participant P6's own words, \textit{``I choose all human features and I think they may have better performance as I know why they were generated.''}
Another notable aspect is the participants' perception of the disparity in the number of AI- and human-suggested features ($t(26) = -5.33, p-value < .01$). Several participants (P4, P7, P8, P12) expressed their concern over the overwhelming number of AI-suggested features presented to them. \textit{`Although the number of features truly mirrors the real situation,  personally, I felt a bit overwhelmed when presented with more than 20 AI features. (P8)''}
Hence, it is essential for human\&AI FE systems to take into account the number of suggested features and devote attention to designing efficient methods for displaying features. This would facilitate users in swiftly exploring the AI-generated features and alleviate their cognitive load. We further discuss this in Sec.~\ref{dis-6.3}.

Overall, participants displayed varying perceptions of features suggested by humans and AI in terms of explainability, complexity, trustworthiness, and scale. Despite these perceived differences, when both human- and AI-generated features were presented together for recommendations, nearly all participants chose features from both sources (Fig.~\ref{fig:result3}. The underlying reasons for this behavior will be elaborated in the subsequent subsection ~\ref{RQ2-complementary}.

\subsubsection{Perceived Complementarity of Features from Human and AI} \label{RQ2-complementary}
Through careful examination and evaluation of various feature combinations during the study, a number of participants (P1, P4, P5, P6, P10, P11, P12) discovered that features from both humans and AI can be \textbf{complementary in terms of domain knowledge coverage}. This observation stems from the recognition that humans possess limited domain knowledge, whereas AI has the capacity to leverage extensive datasets or external sources, such as knowledge graphs, to augment knowledge and generate features.
As P1 mentioned, \textit{``AI-suggested features may fill in the `blind spots' of human knowledge by encompassing a broader scope that I may not have considered in my manual process.''}
Notably, some participants (P1, P5, P10, P11) initially favored human features and subsequently incorporated several AI-generated features to assess if the performance score would improve. 
They regarded human-generated features as the primary component and utilized AI-generated features to enhance overall performance when they realized that relying solely on human-generated features was insufficient.

Furthermore, participants consistently reported that human-generated features are less complex compared to those generated by AI. As a result, some participants (P4, P6, P12) recognized \textbf{the potential of AI-generated features to produce innovative and comprehensive features}. Participant P12 stated, \textit{``It seems that the AI features are more complex and sometimes they can be more creative than humans' features. At least I have limited knowledge and cannot build a complex feature like this.''}
This valuable finding emphasizes the potential of human\&AI FE and the practicality of combining human- and AI-generated features in data science workflows. We further discuss the complementarity of these features in Sec.~\ref{dis-6.2}.

\subsection{\textbf{User Experience of the Human\&AI FE Design (RQ2)}} \label{RQ2-interface}
We also collected participants' general experience and feedback on the design of human\&AI FE, the interface, and the overall workflow (Fig.~\ref{fig:study1}). In general, most participants showed their interest and gave positive feedback on using the workflow ($MEAN = 6.64, SD = 0.50$). As mentioned by participant P8, \textit{``I like a mix of votes and I think showing the creator is useful. I could choose to use one side or analyze both sides to gain insight.''}
Most participants (11 out of 14) felt the interface (i.e., tabular view) for presenting features was friendly to use and interact with ($MEAN = 6.36, SD= 0.49$). Participant P1 noted, \textit{``It is an intuitive way as it seamlessly connects the recommended features with code cells in the familiar setting [online notebooks].''}
All participants appreciated the decision to develop a plugin in Jupyter notebooks, as they often work with online notebooks and share them with others. Additionally, participants highlighted that the simple design and seamless integration with the notebook were beginner-friendly, enabling them to run the workflow with default settings without writing new code.

Participants generally expressed high appreciation for the human\&AI FE design, particularly the intuitive tabular interface. However, they also identified certain drawbacks and provided valuable suggestions for improvement (see Sec.~\ref{RQ3}), which opens up research opportunities for future human\&AI FE systems.

\subsection{\textbf{Design Improvements Requested By the Participants (RQ3)}} \label{RQ3}
We categorized participants' feedback into \textbf{five aspects}:
\textbf{(1)} Requiring context information of features to enhance features' reuse and sharing, 
\textbf{(2)} facilitating complementary feature recommendations from humans and AI, 
\textbf{(3)} supporting flexible and transparent visualizations for human\&AI FE, 
\textbf{(4)} showing the number of features based on the tasks and user needs,
and \textbf{(5)} exploring more adaptable recommendation approaches in data science workflows. 
These suggestions provide design considerations for feature engineering systems design that includes human and AI.

\subsubsection{Requiring context information of features} \label{RQ3-1}
Participants expressed a need for \textbf{more contextual information about AI-generated features.} 
They highlighted the importance of transparency in introducing AI-generated features to enhance trust, including the reasons behind and the generation process. 
For example, two participants (P6 and P11) expressed the need for a metric, such as a recommended score, to guide their exploration of AI-generated features. 
In terms of enhancing feature reuse and sharing, participants highlighted the need for \textbf{social context information.} Specifically, six participants (P1, P5, P6, P11, P12, P13) expressed their desire to access the usage history and comments from others regarding the suggested features. 
For instance, participant P11 recommend \textit{``providing information about other users' selection history, comments, or some context of these features.''}
Two participants (P6, P12) also expressed a desire for information about the \textit{``Creators''} of the features to assess the reliability and usefulness of the suggested features, such as the creators' background, experience, and the motivations behind feature creation. 

\subsubsection{Facilitating complementary feature recommendation} \label{RQ3-3}
Participants expressed strong interest in the complementary part between humans and AI in the FE process (Sec.~\ref{RQ2-complementary}), and provided two practical suggestions.
First, they (P5, P13, P14) recommended \textbf{suggesting AI-generated features alongside human features as supplementary options}, such as \textit{``grouping or aligning AI features with a recommended human feature that shares the same input.''}
This approach allows for easy evaluation and decision-making regarding the use of AI features to enhance and compensate for the limitations of human features. Participants emphasized their tendency to rely on human features initially and consider exploring AI features when human features underperform or when they have limited (domain) knowledge.

Second, participants (P2, P12) highlighted the importance of \textbf{providing human and AI features based on the specific stages of projects}. They noted that practitioners may have different feature needs at various stages of the data science pipeline.
For example, participant P12 said that \textit{``AI features are useful for quick prototyping and dataset evaluation. However, when it comes to interpreting and persuading others, I prefer using human features, even if their performance is slightly inferior.''} 
By incorporating these suggestions, human\&AI FE systems can leverage the strengths of both humans and AI, fostering a more effective and collaborative FE process.

\subsubsection{Supporting flexible and various visualizations for human\&AI FE} \label{RQ3-4}
Participants provided valuable insights on the importance of supporting flexible and transparent visualizations in the human\&AI FE. 
For instance, participant P6 suggested incorporating customized filtering functions, such as filtering features by specific transformers (e.g., $MAX()$ or $MIN()$), as these functions could greatly facilitate feature selection. Additionally, participants P2, P8, and P13 highlighted the need for a more flexible tabular view that can accommodate a larger number of features on a single page. 
In addition to improving the tabular view, participants emphasized the importance of \textbf{providing other forms of visualization} to enhance feature engineering. Several participants (P6, P11, P14) recommended the inclusion of a correlation map, which would display the correlations between human and AI features. Furthermore, participants P1, P5, P9, and P14 suggested the use of clustering techniques to group similar features together, enabling the identification of redundant features and facilitating the identification of useful ones more efficiently.
These suggestions provide valuable insights for enhancing the user interface of future human\&AI FE system, allowing for flexible interaction with features and the utilization of data visualizations to aid in effective FE.

\subsubsection{Exploring adaptable feature recommendation approaches} \label{RQ3-5}
In terms of the feature recommendation, participants (P1, P4, P5, P13, P14) provided suggestions to enhance the user experience of human\&AI FE. One participant (P14) emphasized the importance of allowing users to have control over the recommended features, recognizing that individuals with diverse backgrounds and work experiences may have varying requirements for feature recommendation. \textit{``The recommendation would be welcome for the users with less experience in creating features or limited domain knowledge. More experienced people may rely less on the recommendations, so future human\&AI FE systems should consider adaptable recommendation strategies. (P14)''}

%% file: Sections/06-discussion.tex
\section{Discussion} 
Through the examination of the three research questions, this study has provided valuable insights and detailed findings pertaining to the practice and experience of data science (DS) practitioners with the human\&AI FE design. In this section, we expound upon the significant lessons derived from the user study, engage in a comprehensive discussion surrounding key issues related to the human\&AI FE, and put forth potential design considerations for the application of human\&AI feature recommendations in future research endeavors.

\subsection{Effects of Showing the Creator of Recommended Features} \label{dis-6.1}
Examining the influence of disclosing the source of recommended features constituted a key objective of our user study. Our findings revealed that the originator of recommended features played a critical role in shaping the feature selection and behaviors of DS practitioners. Overall, our results indicated a greater degree of trust placed in human-generated features compared to AI-generated ones. 
Moreover, we observed variations among participants in terms of their reliance and dependence on the \textit{creator} information. These outcomes reinforce the significance of providing \textit{creator} information within the context of human\&AI FE.

Despite its significance, we recommend that designers carefully consider whether and how to disclose the \textit{creator} information when presenting both human- and AI-generated features to DS practitioners.  
On one hand, disclosing this information can enhance transparency and assist users in making informed choices by allowing them to evaluate the source of the features. On the other hand, designers should be cautious about the potential pitfalls of showing such information. Participants' feedback highlighted that prior beliefs and experiencescould influence users' perceptions and reliance, potentially leading to biases in their decision-making process~\cite{vodrahalli2021humans, zhu2022bias}.Specifically, if a user has a strong belief or trust in the human-suggested features, showing the source of features may affect the user's selection as he/she may not fully exploit AI-suggested features. Conversely, if a user has a strong belief in the reliability of AI-generated features, they may overlook the potential value of human-generated features.
As a result, future designs need to correctly guide users' perception of features and help users jump out of the prejudice against \textit{creators}. Moreover, more empirical investigations (e.g., controlled experiments) are needed to deeply understand the effects of showing the \textit{creator} of features on users' perceptions in various tasks.


\subsection{Complementarity of Human and AI Assistance in Feature Engineering} \label{dis-6.2}
Our qualitative findings revealed that participants recognized the complementary nature of features from both human and AI sources, particularly in terms of data attribute coverage, feature quantity, complexity, and interpretability. Participants acknowledged that AI-generated and human-generated features can contribute to different aspects and stages of the feature engineering (FE) process. This observation is further supported by the final feature selections made by participants, as depicted in Fig.~\ref{fig:result3}, where the majority chose features from both human and AI sources for their final feature set.
While existing research often focuses on comparing the similarities and differences between AI and humans~\cite{vodrahalli2021humans, biessmann2021turing}, with some aiming to make AI assistance emulate human behavior~\cite{ma2022glancee, ma2019smarteye}, there is a scarcity of work specifically examining the complementarity between the two~\cite{bansal2021does}.
However, we believe that investigating how humans and AI can effectively complement each other is a valuable direction for future research in human-AI FE. Our goal is to surpass the capabilities of individual practitioners or AI alone, rather than fully automating the FE process.

Therefore, our researchers could investigate how to enhance the complementarity of human and AI through human-centered design. 
In addition, research on human-AI collaboration in data science offered many insights into how to combine and leverage the benefits of both sides from a collaboration view ~\cite{wang2019human, automationsurvey}, which can also be considered in the feature recommendation scenario.
To achieve complementariry, we need to understand the differences between human and AI assistance and analyze the advantages and disadvantages of both human-and AI-generated features before we apply them in feature recommendation systems. Then, we can try to make the complementarity more explicit. 
For example, filtering out AI-recommended features that are similar to human-generated ones while retaining distinct contributions can enhance complementarity. Additionally, training AI models with the objective of complementing human partners can be explored~\cite{bansal2021most}. By identifying the specific needs and limitations of DS practitioners and redesigning AI models accordingly, we can address tasks that humans struggle with. Through these efforts, we can foster more effective and collaborative feature engineering practices.


\subsection{Transparency of Feature Recommendation} \label{dis-6.3}
Transparency issues surrounding feature recommendation, particularly with AI-generated features, were identified in our study. These issues can be categorized into two aspects. 
First, DS practitioners expressed concerns about the lack of explainability of AI-suggested features, either due to complex compositions or incomprehensible descriptions/definitions (see Section ~\ref{RQ2-complementary}). 
The lack of interpretability hinders the adoption of these features, leading to \textit{under-trust}~\cite{lee2004trust}, where effective features are disregarded due to a lack of interpretability. To address this, improvements should be made in enhancing the interpretability of features. Advanced natural language generation techniques (e.g., GPT-4) can be employed to improve the readability of feature descriptions and enhance understanding~\cite{hollmann2023gpt}.


Second, participants expressed a desire for context information about the recommended features, as it enhances transparency. 
The context information includes: which source attributes a feature comes from, how a feature is generated, why it is recommended, where a feature is distributed in the semantic space, etc. 
However, it is important to strike a balance as displaying excessive information can overwhelm users~\cite{ma2022glancee}. 
One approach is to utilize visualizations that present information in a more comprehensible manner~\cite{collaris2020explainexplore, chatzimparmpas2021featureenvi, rojo2020gacovi}. Another approach is to provide context information on-demand, triggered and displayed only when the user requires clarification on a specific recommended feature. 
In general, what information should be transparent and how to display the additional context of recommended features are deserved to explore in the future, and the detailed setting or design can be obtained from further analysis of user requirements in specific context or tasks.

Third, improving social transparency (ST) ~\cite{ehsan2021expanding} is a viable method for enhancing human\&AI FE. For instance, disclosing whether other users accepted or rejected AI recommendations can provide valuable context in AI recommendation systems~\cite{ehsan2021expanding}.
Additionally, providing information on how features are used by other DS practitioners and contributor-related information for human-generated features can aid in better decision-making. 
In summary, designing transparent feature recommendations should be driven by user needs and carried out through human-centered design. Empirical studies are crucial to verify the effectiveness of transparency.

\subsection{Personalized and Adaptive Feature Recommendation} \label{dis-6.4}
Our user study revealed that participants employed varied strategies and displayed diverse preferences in the feature selection process~\ref{strategy-creator}, likely influenced by their individual knowledge and prior work experience in FE. To accommodate these differences, interface design should cater to users' unique FE habits and needs. For instance, a configuration page could be provided, enabling users to select their preferred feature recommendation methods and related support functions. 
Additionally, monitoring the code or edit history in the notebook would allow for timely feature suggestions when needed. For example, citizen DS practitioners, who tend to focus on one to two feature categories (see Sec.\ref{RQ1-4}), could benefit from reminders to consider other features, enhancing feature diversity and model performance.

Furthermore, designers can leverage user modeling approaches, employing interactive machine learning methods, to capture users' preferences~\cite{ma2019smarteye, amershi2014power, ye2024contemporary}. For instance, offering functions that enable users to provide feedback on the usefulness of recommended features, akin to the ``like'' or ``dislike'' options in recommendation systems, can facilitate learning users' feature selection preferences. By incorporating this feedback loop, the feature recommendation model can continually refine its recommendations, ultimately improving recommendation quality.


In summary, personalization and adaptation based on individual experience and preference are essential to the success of human\&AI FE, especially when the feature recommendation becomes a daily assistive tool for DS practitioners in the future.

\subsection{Limitation and Future Work} \label{dis-6.5}
This work represents an initial step towards exploring the integration of human- and AI-generated features in the FE stage of the data science lifecycle. However, there are certain limitations that need to be acknowledged. 
First, our evaluation of the human-AI FE design probe was conducted using a single case study~\cite{rosson2009scenario}, which may not capture the full range of user behaviors and specific needs that could arise in different scenarios. 
While we ensured protocol consistency among participants and captured common user behaviors and needs in the selected case, it is important to recognize that our study should be viewed as formative rather than evaluative. Future work should involve conducting additional empirical evaluations to further enhance our understanding of how DS practitioners interact with and perceive such a human-AI collaborative FE system.
Second, although we anticipated that AI would generate a larger number of FE recommendations compared to humans (useful or not is a different story), we observed an unbalanced distribution of human- and AI-generated features with the human\&AI FE in our exploratory study. Specifically, the number of human-generated features was significantly lower than AI-generated features. This disparity could potentially influence participants' feature selection outcomes. For instance, if participants were provided with more human-generated features, they might rely less on AI-generated features. Therefore, in future studies, it would be valuable to explore the impact of the quantity of AI-recommended and human-recommended features on users' perceptions and adoption of recommendations.
Third, our findings may only be applicable to scenarios involving presenting tabular data to DS practitioners, and creating or applying features with other types of data representation (e.g., hierarchical or spatial data) may require further validation.


Furthermore, the proposed human\&AI FE should be tested with a wider range of stakeholders and applied to various feature engineering tasks. Additionally, a long-term study is also necessary to observe whether DS practitioners' behavior patterns and perceptions of human\&AI feature recommendations could get shaped by the increasing collaboration with the system. 
Overall, while this work presents a technically feasible exploration of combining human and AI feature recommendations, it serves as a starting point for pushing the boundaries of human-AI integration designs. 

%% file: Sections/07-conclusion.tex
\section{Conclusion}
This paper investigates data science practitioners' perceptions and utilization of feature recommendations in the context of human-AI collaboration. We propose and implement a feature recommendation prototype that integrates both human and AI assistance. A user study involving 14 participants is conducted to gain insights into their usage patterns and perceptions of feature recommendations from humans and AI. The results emphasize the benefits of combining human and AI assistance in feature engineering. Practical implications for future human-AI collaborative design are discussed, with the aim of advancing the understanding and effectiveness of human-AI collaboration in feature engineering.